\documentclass[twocolumn, pra, showpacs,superscriptaddress]{revtex4}
\usepackage{amssymb}
\usepackage{graphicx}
\usepackage{dcolumn}
\usepackage{bm}
\usepackage{amsmath}

\begin{document}

\title{Density-functional theory for 1D harmonically trapped Bose-Fermi
mixture}
\author{Hongmei Wang}
\affiliation{Department of Physics and Institute of Theoretical Physics, Shanxi
University, Taiyuan 030006, P. R. China}
\affiliation{Department of Physics, Taiyuan Normal University, Taiyuan 030001, P. R. China}
\author{Yajiang Hao}
\affiliation{Department of Physics, University of Science and Technology Beijing, Beijing 100083, P. R.
China}
\author{Yunbo Zhang}
\email{ybzhang@sxu.edu.cn}
\affiliation{Department of Physics and Institute of Theoretical Physics, Shanxi
University, Taiyuan 030006, P. R. China}

\begin{abstract}
We present a density-functional theory for the one dimensional
harmonically trapped Bose-Fermi mixture with repulsive contact
interactions. The ground state density distribution of each
component is obtained by solving the Kohn-Sham equations numerically
based on the Local Density Approximation and the exact
solution for the homogeneous system given by Bethe ansatz method.
It is shown that for strong enough interaction, a considerable amount of fermions are
repelled out of the central region of the trap, exhibiting partial phase
separation of Bose and Fermi components. Oscillations emerge in the 
Bose density curves reflecting the 
strong correlation with Fermions. For infinite strong interaction, 
the ground state energy of the mixture and the total density 
are consistent with the scenario that all atoms in the mixture 
are fully fermionized. 

\end{abstract}

\maketitle

\section{Introduction}

Ultracold atomic gases provide a highly controllable testing ground
to study fundamental problems in quantum many body physics
\cite{Pethick} and many experimental observations can be compared 
directly with exactly solvable theories. The degenerate quantum gases with
multi-component in low spatial dimensions, especially in one dimension (1D), 
become one of the growing interesting topics \cite{CazalillaRMP}.
Multi-component gases can be mixture of the same species of atoms with
different hyperfine states, i.e. spinor condensate, or mixture of different
species of atoms. The competition between the inter- and intra-species
interaction makes the mixture system more complicated and there exhibit
richer physical phenomena than its single-component counterpart. 
Bose-Fermi mixture is originally
realized in experiments as a result of sympathetic cooling technique, i.e.,
cooling the fermions to quantum degeneracy through the mediation of bosons
\cite{BFM experiment 1, BFM experiment LiLi, BFM experiment K Rb, BFM
experiment YbYb, BFM experiment KK}. Then many theoretical studies have been
performed on three dimensional mixture, dealing with the phase 
separation \cite{Molmer}, 
pairing \cite{Lewenstein2004}, superfluids and Mott insulators transition 
\cite{Hebert2007}, BEC and Bardeen-Cooper-Schrieffer crossover 
\cite{Illuminati2004}, etc. On the other hand, 1D systems 
attract attentions for the simplicity of theoretical models and for the significance
of quantum correlation effects therein \cite{1D book}. Experimentally the 1D 
systems can be realized by confining the cold atoms in two dimensional optical 
lattices or in strong anisotropic magnetic trap \cite{1D experiment}. The interaction
among the atoms can be tuned in the whole regime of interaction strength via
the magnetic Feshbach resonance and controlling
the transverse confinement of magnetic trap \cite{Olshanii}. Interestingly
enough, the properties of these system, including the ground state, 
elementary excitations as well as thermodynamics, are sometimes fairly well 
captured by the exactly solvable models studied a few decades ago \cite{Sutherland}.

Many theories have studied the 1D Bose-Fermi mixture both homogeneous and 
trapped gas in external potentials. When the system is homogeneous, 
Das \cite{Das}
early plots the ground state phase diagram based on the mean-field theory, which
predicts the occurrence of phase separation (i.e., demixing) of the two
components. Luttinger liquid formalism shows that for strong enough repulsion 
the two components of the mixture, with bosons either a quasicondensate or 
impenetrable particles, repel each other sufficiently to demix \cite{Cazalilla2003}.
But the exact Bethe ansatz solution for the 1D mixture with
equal mass and equal coupling constants points out the absence of demixing \cite%
{LaiYang,Imambekov}. It indicates that mean-field theory
and Luttinger liquid theory are reliable only for very weak interaction. 
With the system loaded in optical lattices, the phase diagram and correlation 
functions have been investigated with the bosonization method \cite{Bosonization} 
and quantum Monte Carlo (MC) simulation \cite{MC}. 
For the trapped system, the local density approximation (LDA) on the Bethe
ansatz solution shows that the harmonic trap 1D mixture would partially demix
for strong repulsive interaction \cite{Imambekov}. The finite temperature 
Yang-Yang thermodynamics and the quantum criticality analysis based on 
thermodynamic Bethe ansatz (TBA) both support the description of phase separation
\cite{Xiangguo Yin}.
In the infinitely strong
interaction limit, i.e. the Tonks-Girardeau (TG) regime, Bose-Fermi mapping method 
\cite{Girardeau2007} gives a result that the density profiles display
no demixing among the two component, where the exact ground state is highly 
degenerate and the most symmetrical one is chosen 
\cite{Bess Fang2009,Buljan2009,Xiaolong2010}. 
Later detailed calculations are done for all 
degenerate manifolds of the ground state \cite{Bess Fang2011} and the 
conclusion of nondemixing remains for the mixture in TG limit.

So far a method is not available for 1D trapped Bose-Fermi mixture suitable for the
whole repulsive interaction regime. Numerical simulations such as Density Matrix 
Renormalization Group (DMRG) and quantum MC are however limited to 
few atom numbers in lattice models \cite{Bess Fang2011,MC}.
In this paper, we develop Hohenberg-Kohn-Sham Density-functional theory
(DFT) to investigate the ground state properties of 1D harmonically trapped
Bose-Fermi mixture. It is well known that DFT is a successful and widely
used approach for treating the electron systems with long range Columb
interaction \cite{Kohn, DFT book}. Recently it has been successfully generalized
to cold atom systems with short-range contact interaction
for three dimensional bosonic atoms \cite{DFT for 3D bosonic systems,Kim2003} and three
dimensional Bose-Fermi mixture \cite{DFT for 3D bose fermion mixture}. In the
framework of DFT, in order to investigate the ground state properties of an
inhomogeneous interacting system, a homogeneous interacting system is often
needed in the process of local density approximation (LDA) for the exchange
correlation energy \cite{Kohn}. Because Bethe ansatz method can give exact
solution for 1D homogeneous systems, several authors developed the DFT based
on Bethe ansatz results to solve the 1D Bosons \cite{Kim2003,
Brand2004, Yajiang2009} and 1D Fermi cold atom systems \cite{DFT for 1D
fermion, Gao xianlong}. Here we apply this method for 1D Bose-Fermion
mixture. As can be seen below, the key points of our scheme include a suitable
fitting formula for the ground state energy and appropriate choices of the 
functional orbitals for boson and fermions. 

The paper is organized as follows. In Sec. II we derive the universal
Kohn-Sham equations with LDA for 1D Bose-Fermi mixture and then present
the expression of exchange-correlation energy. From the exact result of Bethe
ansatz for homogeneous gas with equal masses of atoms and equal interaction 
of boson-bosn and boson-fermion we find a fitting formula for the ground state
energy to simplify the numerical iterations. Then the equations
are solved numerically and the ground state energy and density distribution
are discussed in Sec. III. Finally we conclude our result in the last section.

\section{Theory}

\subsection{Kohn-Sham equations}

We consider a 1D trapped mixture of $N_{B}$ bosons and $N_{F}$
spin-polarized fermions with two-body contact interactions.
$N=N_{B}+N_{F}$ is the total atom number. The system is described by the Hamiltonian%
\begin{eqnarray}
H &=&\sum_{i=1}^{N_{B}}\left[ -\frac{\hbar ^{2}}{2m_{B}}\frac{d^{2}}{%
dx_{i}^{2}}+V_{B}(x_{i})\right]  \notag \\
&&+\sum_{j=1}^{N_{F}}\left[ -\frac{\hbar ^{2}}{2m_{F}}\frac{d^{2}}{dx_{j}^{2}%
}+V_{F}(x_{j})\right]  \notag \\
&&+\frac{g_{BB}}{2}\sum_{i,i^{\prime }=1}^{N_{B}}\delta \left(
x_{i}-x_{i^{\prime }}\right)  \notag \\
&&+g_{BF}\sum_{i=1}^{N_{B}}\sum_{j=1}^{N_{F}}\delta \left(
x_{i}-x_{j}\right) .  \label{H}
\end{eqnarray}%
Here $m_{B}$, $m_{F}$ are boson and fermion masses, $V_{B}(x)$, $V_{F}(x)$
are external potentials, and $g_{BB}$, $g_{BF}$ are the effective 1D Bose-Bose
and Bose-Fermi interaction parameters, which can be tuned experimentally%
\cite{Olshanii,Rizzi2008}. The Fermi-Fermi interaction is
not considered because the Pauli exclusion principle suppresses the contact $%
s$-wave scattering and their $p$-wave scattering can be neglected.

According to the Hohenberg-Kohn theorem I of DFT \cite{Kohn}, the ground
state density of a bound system of interacting particles in some external
potential determines this potential uniquely. It thus gives us the full
Hamiltonian (\ref{H}) and particle number $N$. Hence the density determines
implicitly all properties derivable from $H$ through the solution of the
time-independent or time-dependent Schr\"{o}dinger equation. Though proved
originally for Fermions, the theorem can be straightforwardly generalized to
Bosons as well as the mixture of Bosons and Fermions studied here. Denote
the densities of bosons and fermions as $n_{B}\left( x\right) $ and $%
n_{F}\left( x\right) $, respectively, the total density is then obviously $%
n\left( x\right) =n_{B}\left( x\right) +n_{F}\left( x\right) $. The number
of Bosons and Fermions are conserved separately, i.e. $\int n_{B}\left(
x\right) dx=N_{B}$, $\int n_{F}\left( x\right) dx=N_{F}$ and $\int n\left(
x\right) dx=N$. The ground state energy, defined as $\left\langle
g\right\vert H\left\vert g\right\rangle $ with $\left\vert g\right\rangle $
the ground state of system, is a functional of the densities $E_{0}\left[
n_{B}\left( x\right) ,n_{F}\left( x\right) \right] $, which can be
decomposed as
\begin{eqnarray}
E_{0} &=&T_{B}^{ref}\left[ n_{B},n_{F}\right] +T_{F}^{ref}\left[ n_{B},n_{F}%
\right]  \notag \\
&&+\int dxn_{B}\left( x\right) V_{B}\left( x\right) +\int dxn_{F}\left(
x\right) V_{F}\left( x\right)  \notag \\
&&+\frac{g_{BB}}{2}\int dxn_{B}^{2}\left( x\right) +g_{BF}\int dxn_{B}\left(
x\right) n_{F}\left( x\right)  \notag \\
&&+E_{xc}\left[ n_{B},n_{F}\right] .  \label{E0}
\end{eqnarray}%
The first two terms are Bose and Fermi kinetic energies of a reference
noninteracting system. The next two terms in the second row are external
potential energies and those in the third row are Hartree-Fock energies
(i.e., the mean-field approximation of the interaction energy). The last
term is the exchange correlation energy which includes all the contributions to
the interaction energy beyond mean-field theory.

We assume the Bosons are in a quasi-condensate state and Fermions are in
normal state. Thus we introduce a single Bose functional orbital $\phi
\left( x\right) $ and $N_{F}$ Fermi functional orbitals $\psi _{j}\left(
x\right) $ ($j=1\cdots N_{F}$) which are orthogonal and normalized. This is
different from the way of Ref. \cite{Adhikari2010} where only one condensed
orbital of fermionic pair is considered for the mixture of bosons and paired
two-component fermions in superfluid state or BCS state. With $\phi \left(
x\right) $ and $\psi _{j}\left( x\right) $, the densities are expressed as
\begin{eqnarray}
n_{B}\left( x\right) &=&N_{B}\phi ^{\ast }\left( x\right) \phi \left(
x\right) ,  \notag \\
n_{F}\left( x\right) &=&\sum_{j=1}^{N_{F}}\psi _{j}^{\ast }\left( x\right)
\psi _{j}\left( x\right) .  \label{n}
\end{eqnarray}%
and the kinetic energies are
\begin{eqnarray}
T_{B}^{ref} &=&-N_{B}\int dx\phi ^{\ast }\left( x\right) \frac{\hbar ^{2}}{%
2m_{B}}\frac{d^{2}}{dx^{2}}\phi \left( x\right) ,  \label{TB} \\
T_{F}^{ref} &=&-\sum_{j=1}^{N_{F}}\int dx\psi _{j}^{\ast }\left( x\right)
\frac{\hbar ^{2}}{2m_{F}}\frac{d^{2}}{dx^{2}}\psi _{j}\left( x\right) .
\label{TF}
\end{eqnarray}%
As far as the exchange correlation energy $E_{xc}\left[ n_{B},n_{F}\right] $
is concerned, when the confinement is weak, we adopt the Local Density
Approximation (LDA), i.e., the system can be assumed locally homogeneous at
each point $x$ in the external trap. In this way $E_{xc}$ is approximated
with an integral over the exchange-correlation energy per atom of a
homogeneous interacting mixture $\varepsilon _{xc}^{\hom }\left(
n_{B},n_{F}\right) $
\begin{equation}
E_{xc}\approx \int dxn\left( x\right) \varepsilon _{xc}^{\hom }\left(
n_{B},n_{F}\right) ,  \label{E xc}
\end{equation}%
where the densities $n_{B}$, $n_{F}$ are taken at point $x$.

For such a homogeneous interacting mixture,
\begin{equation}
\varepsilon _{xc}^{\hom }=\varepsilon ^{\hom }-\varepsilon _{M}^{\hom
}-\kappa _{s}^{\hom },  \label{e xc}
\end{equation}%
where $\varepsilon ^{\hom }$ is the ground state energy per atom; $%
\varepsilon _{M}^{\hom }=g_{BB}n_{B}^{2}/2n+g_{BF}n_{B}n_{F}/n$ is the mean
field interaction energy per atom; $\kappa _{s}^{\hom }=\hbar ^{2}\pi
^{2}n_{F}^{3}/6m_{F}n$ is the kinetic pressure terms, i.e., the total
kinetic energy dividing by the total number of fermions and bosons in a
noninteracting homogeneous mixture. Here the kinetic energy of the bosons is
easily shown to be zero and the kinetic energy comes solely from the
exclusive quantum state occupation of fermions .

Hohenberg-Kohn theorem II \cite{Kohn} guarantees that the ground state
density distributions is determined by variationally minimizing $E_{0}$ with
respect to $n_{B}\left( x\right) $ and $n_{F}\left( x\right) $, which is
equivalent to a variational calculation of (\ref{E0}) with respect to the Bose
and Fermi functional orbitals $\phi ^{\ast },\psi _{j}^{\ast }$. After
substituting (\ref{e xc}) into (\ref{E xc}) and substituting (\ref{n})-(\ref%
{E xc}) into (\ref{E0}), we carry out the functional derivatives
\begin{eqnarray}
\left. \delta \left( E_{0}-\epsilon N_{B}(\int \phi ^{\ast }dx\phi
-1)\right) \right/ \delta \phi ^{\ast } &=&0,  \notag \\
\left. \delta \left( E_{0}-\sum_{j=1}^{N_{F}}\eta _{j}(\int \psi _{j}^{\ast
}dx\psi _{j}-1)\right) \right/ \delta \psi _{j}^{\ast } &=&0,
\label{derivative}
\end{eqnarray}%
where $\epsilon $ and $\eta _{j}$ ($j=1,2,\cdots N_{F}$) are Lagrange
multipliers conserving the normalization of $\phi \left( x\right) $ and $%
\psi _{j}\left( x\right) $. Then we can get the Kohn-Sham equations (KSEs)%
\begin{widetext}
\begin{eqnarray}
\left( -\frac{\hbar ^{2}}{2m_{B}}\frac{d^{2}}{dx^{2}}+V_{B}\left( x\right)
+\mu _{B}^{\hom }\left( \left[ n_{B},n_{F}\right] ;x\right) \right) \phi
\left( x\right)  &=&\epsilon \phi \left( x\right) ,  \label{KSB} \\
\left( -\frac{\hbar ^{2}}{2m_{F}}\frac{d^{2}}{dx^{2}}+V_{F}\left( x\right) -%
\frac{\hbar ^{2}}{2m_{F}}\pi ^{2}n_{F}^{2}\left( x\right) +\mu _{F}^{\hom
}\left( \left[ n_{B},n_{F}\right] ;x\right) \right) \psi _{j}\left( x\right)
&=&\eta _{j}\psi _{j}\left( x\right) .  \label{KSF}
\end{eqnarray}%
\end{widetext}Here $\mu _{B}^{\hom }=\partial \left( n\varepsilon ^{\hom
}\right) /\partial n_{B}$ and $\mu _{F}^{\hom }=\partial \left( n\varepsilon
^{\hom }\right) /\partial n_{F}$ are Bose and Fermi chemical potentials of a
homogeneous interacting mixture. Physically $\epsilon $ and $\eta _{j}$ are
the lowest eigenvalues of KSE. In (\ref{n}), the sum in $n_{F}\left(
x\right) $ runs over the occupied orbitals $\psi _{j}$ with lowest $\eta
_{j} $.

Left multiplying $\psi _{j}^{\ast }$ on both sides of (\ref{KSF}), performing 
summation over $j$ and integrating over $x$, we get an expression of 
$T_{F}^{ref}$ defined in (\ref{TF}). Analogously from the
normalization of $\phi (x)$ we may get an expression for $T_{B}^{ref}$ defined
in (\ref{TB}). Inserting these two kinetic terms into (\ref{E0}), the ground
state energy (\ref{E0}) is expressed as a function of $\epsilon $ and $\eta _{j}$
\begin{eqnarray}
E_{0} &=&N_{B}\epsilon +\sum_{j=1}^{N_{F}}\eta _{j}  \notag \\
&&+\int n\left( x\right) \varepsilon ^{\hom }\left( x\right) dx-\int
n_{B}\left( x\right) \mu _{B}^{\hom }\left( x\right) dx  \notag \\
&&-\int n_{F}\left( x\right) \mu _{F}^{\hom }\left( x\right) dx+\frac{\hbar
^{2}\pi ^{2}}{3m_{F}}\int n_{F}^{3}\left( x\right) dx  \label{E0c}
\end{eqnarray}

If $\varepsilon ^{\hom }\left[ n_{B},n_{F}\right] $ are known, we can solve
the KSEs together with (\ref{n}) to find the density distributions $%
n_{B}\left( x\right) $, $n_{F}\left( x\right) $ and then calculate the
ground state energy $E_{0}$ from (\ref{E0c}). In the following
we present the result of $\varepsilon ^{\hom }\left[ n_{B},n_{F}\right] $ by
means of the Bethe ansatz method.

\subsection{Ground state energy of homogeneous system}

In the absence of an external trap the system is homogeneous, which can be
solved exactly via Bethe ansatz method for a much restrictive but simple case%
\begin{eqnarray}
g_{BB} &=&g_{BF}=g>0,  \notag \\
m_{B} &=&m_{F}=m.  \label{mB=mF}
\end{eqnarray}%
It describes the situation that the interactions of Boson-Boson and
Boson-Fermion are repulsive with the same strength, and the masses of Boson
and Fermion are the same too. Detailed possible ways to realize this
situation in cold atom experiments have been considered previously \cite%
{Imambekov}. The first condition can be satisfied using the
combination of Feshbach resonance (to control the interactions) and
appropriate choice of the tuning of the trapping laser frequencies (to
adjust the the ratio of the radial confinement of Bosons and Fermions). The
second condition is approximately satisfied with a mixture of two isotopes
of a species of atoms. Isotope mixture is widely used in experiment for it
can avoid the gravitational sag of an external potential caused by different
masses. The experiments have realized 3D isotope mixtures $^{6}$Li-$^{7}$Li
\cite{BFM experiment LiLi}, $^{173}$Yb-$^{174}$Yb \cite{BFM experiment YbYb}%
, $^{40}$K-$^{41}$K \cite{BFM experiment KK} and we see no obvious obstacles
in 1D. Under these two conditions, the Hamiltonian of 1D homogeneous
Bose-Fermi mixture is%
\begin{equation}
H=-\frac{\hbar ^{2}}{2m}\sum_{i=1}^{N}\frac{d^{2}}{dx_{i}^{2}}%
+g\sum_{i<j}\delta \left( x_{i}-x_{j}\right) \text{. }  \label{H-2}
\end{equation}%
This model is solved by means of Bethe ansatz method by Lai and Yang in 1971
for the 1D mixture of bosons and spin-1/2 fermions \cite{LaiYang}. 
Imambekov and Demler investigated the ground state properties in detail
for the 1D mixture of bosons and spin-polarized fermions \cite{Imambekov}
and extensive studies have been done in \cite{conformal quantum
field theory, Guan2008, Xiangguo Yin, Hao2011}
including the thermodynamics and correlation functions. Here we briefly
review the main results of \cite{LaiYang,Imambekov, Hao2011} 
which are readily used as the homogeneous reference system
in our DFT theory. Under the periodic boundary condition and in the
thermodynamical limit (the system size and the number of atoms are
infinitely large but the atomic densities are kept finite), the ground state
Bethe ansatz integral equations are%
\begin{align}
\rho \left( k\right) & =\frac{1}{2\pi }\left[ 1+\int_{-B}^{B}\frac{c\sigma
\left( \Lambda \right) d\Lambda }{c^{2}/4+\left( \Lambda -k\right) ^{2}}%
\right] ,  \notag \\
\sigma \left( \Lambda \right) & =\frac{1}{2\pi }\int_{-Q}^{Q}\frac{c\rho
\left( k\right) dk}{c^{2}/4+\left( \Lambda -k\right) ^{2}},  \label{Bae1}
\end{align}%
where $c=mg/\hbar ^{2}$, $k$\ and $\Lambda $\ are the quasi-momenta and
spectral parameters and $\rho \left( k\right) $\ and $\sigma \left( \Lambda
\right) $\ denote their corresponding density distributions. The integration
limits $B$ and $Q$ are determined by the normalization condition%
\begin{align}
n_{B}& =\int_{-B}^{B}\sigma \left( \Lambda \right) d\Lambda ,  \notag \\
n& =\int_{-Q}^{Q}\rho \left( k\right) dk.  \label{nor1}
\end{align}%
The ground state energy per atom is written in our notation as%
\begin{equation}
\varepsilon ^{\hom }\left( n_{B},n_{F},g\right) =\frac{1}{n}\int_{-Q}^{Q}%
\frac{\hbar ^{2}k^{2}}{2m}\rho \left( k\right) dk.  \label{energy 1}
\end{equation}

For convenience, let us define the fraction of Bosons $\alpha =n_{B}/n$ and
the dimensionless Lieb-Liniger parameter $\gamma =mg/\left( \hbar
^{2}n\right) $. We then introduce variables $x=k/Q$ and $y=\Lambda /B$ such
that $\rho \left( k\right) =\rho \left( xQ\right) =g_{c}\left( x\right) $
and $\sigma \left( \Lambda \right) =\sigma \left( yB\right) =g_{s}\left(
y\right) $, and (\ref{Bae1})-(\ref{energy 1}) are transformed into
\begin{align}
g_{c}\left( x\right) & =\frac{1}{2\pi }\left[ 1+\frac{1}{\lambda _{s}}%
\int_{-1}^{1}\frac{g_{s}\left( y\right) dy}{1/4+\left( y/\lambda
_{s}-x/\lambda _{c}\right) ^{2}}\right] ,  \notag \\
\text{ }g_{s}\left( y\right) & =\frac{1}{2\pi }\frac{1}{\lambda _{c}}%
\int_{-1}^{1}\frac{g_{c}\left( x\right) dx}{1/4+\left( y/\lambda
_{s}-x/\lambda _{c}\right) ^{2}},  \label{Bae2}
\end{align}%
with
\begin{eqnarray}
\lambda _{c} &=&\gamma \int_{-1}^{1}g_{c}\left( x\right) dx,  \notag \\
\lambda _{s} &=&\frac{\gamma }{\alpha }\int_{-1}^{1}g_{s}\left( y\right) dy,
\label{nor2}
\end{eqnarray}%
and
\begin{equation}
\varepsilon ^{\hom }\left( n,\gamma ,\alpha \right) =\frac{\hbar ^{2}n^{2}}{%
2m}e\left( \gamma ,\alpha \right) .  \label{energy 2}
\end{equation}
Here the function
\begin{equation}
e\left( \gamma ,\alpha \right) =\frac{\gamma ^{3}}{\lambda _{c}^{3}}%
\int_{-1}^{1}x^{2}g_{c}\left( x\right) dx  \label{e}
\end{equation}%
can be solved numerically with the combination of (\ref{Bae2}) and (\ref%
{nor2}) by the iteration method. In the limiting cases of $\alpha =0,1$,
the system is purely fermions or purely bosons. $e\left( \gamma ,0\right)
=\pi ^{2}/3$ is a constant while $e\left( \gamma ,1\right) $ coincides with $%
e_{L-L}\left( \gamma \right) $ in the Lieb-Liniger model \cite{Lieb1963}.
When the interaction is weak, $\gamma \ll 1$, the mean field result of 
(\ref{energy 2}) is already available in \cite{Das,Imambekov};
when the interaction is strong, $\gamma \gg 1$, one can neglect the 
dependence of the first integrand in (\ref{Bae2}) on $x$ and $g_{c}\left( x\right) $ 
can be approximated as a constant $g_{c}$. Therefore we get the asymptotic 
behavior of the function $e\left( \gamma ,\alpha \right) $
for $\gamma $
\begin{eqnarray}
e\left( \gamma \rightarrow 0,\alpha \right) &=&\frac{\pi ^{2}}{3}\left(
1-\alpha \right) ^{3}+\left( 2\alpha -\alpha ^{2}\right) \gamma ,  \notag \\
e\left( \gamma \rightarrow +\infty ,\alpha \right) &=&\frac{\pi ^{2}}{3}%
\left( 1-\frac{4F\left( \alpha \right) }{\gamma }+\frac{12F^{2}\left( \alpha
\right) }{\gamma ^{2}}\right) ,  \label{elimit}
\end{eqnarray}%
where $F\left( \alpha \right) =\alpha +\sin \left(\alpha \pi \right)/\pi $. In the
limiting case of $\gamma =0$, $e\left( 0,\alpha \right) =\pi
^{2}\left( 1-\alpha \right) ^{3}/3$, therefore $\varepsilon ^{\hom }\left(
n,0,\alpha \right) =\hbar ^{2}\pi ^{2}n_{F}^{3}/6mn =\kappa
_{s}^{\hom }$, the energy comes solely from the kinetic energy of free
fermions. In the Tonks-Girardeau limit, $e\left( +\infty ,\alpha \right) =\pi ^{2}/3$, which means the
energy of Bose-Fermi mixture with infinitely strong repulsive interactions
is equal to the energy of all atoms treated as free fermions.

\begin{figure}[tbp]
\includegraphics[width=0.45\textwidth]{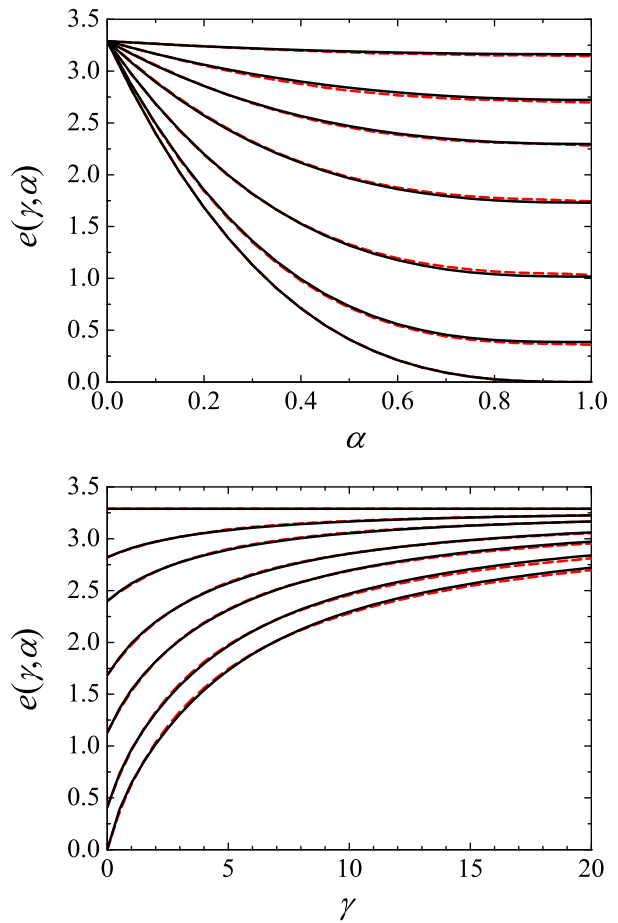}
\caption{ (Color Online) The function $e\left( \protect\gamma ,\protect\alpha \right) $ for
the ground state energy of homogeneous Bose-Fermi mixture system. Numerical
exact result (red dashed line), obtained from the solution of (\protect\ref%
{Bae2}) and (\protect\ref{nor2}) is compared with the fitting formula (solid
black line) given by (\protect\ref{fited e}). (a) $\protect\gamma $%
=0,0.5,2,5,10,20,100 from bottom to top; (b) $\protect\alpha $%
=0,0.05,0.1,0.2,0.3,0.5,1 from top to bottom.}
\label{comparee}
\end{figure}

For practical use, we need calculate $e\left( \gamma ,\alpha \right) $ for
a lot of points $\left( \gamma ,\alpha \right) $. If we use numeric
iteration method for every point, it will be very time-consuming. To avoid
this, we managed to retrieve a parametrization formula for 
$e\left( \gamma ,\alpha \right) $ based on the above limiting cases, which reads as
\begin{eqnarray}
\tilde{e}\left( \gamma ,\alpha \right) &=&\frac{\pi ^{2}}{3}\left( 1-\alpha
\right) ^{3}  \notag \\
&&+f_{1}\left( \gamma \right) (1+f_{2}\left( \gamma \right) \left( 1-\alpha
\right) ^{2}  \notag \\
&&-\left( 1+f_{2}\left( \gamma \right) \right) \left( 1-\alpha \right) ^{3}).
\label{fited e}
\end{eqnarray}%
Here $f_{1}\left( \gamma \right) $ is the approximation of $e_{L-L}\left(
\gamma \right) $. We give
\begin{equation}
f_{1}\left( \gamma \right) =\frac{\pi ^{2}}{3}\frac{\gamma ^{3}+a_{2}\gamma
^{2}+a_{1}\gamma }{\gamma ^{3}+b_{2}\gamma ^{2}+b_{1}\gamma +b_{0}}
\end{equation}%
with $b_{1}=11.37$, $b_{2}=4.68$, $a_{1}=12+b_{1}-4b_{2}$, $a_{2}=-4+b_{2}$
and $b_{0}=\pi ^{2}a_{1}/3$, which exhibits the same asymptotic behavior 
as $e_{L-L}\left(\gamma \right) $ in
the weak and strong interaction cases to the order of $\gamma $ and $%
1/\gamma ^{2}$, respectively. The function $f_{2}\left( \gamma \right) $ is determined
by the numerical iteration result for some sampled values of $\gamma $, and we fit it as
\begin{equation}
f_{2}\left( \gamma \right) =c_{0}\exp (c_{1}\gamma )-\left( c_{0}+1\right)
\exp (c_{2}\gamma )
\end{equation}%
with $c_{0}=0.21$, $c_{1}=-0.02$, $c_{2}=-1.45$. $\tilde{e}\left( \gamma
,\alpha \right) $ gives the exact behavior at the limits $\alpha =0,\gamma
=0,+\infty $ and approximates to $e_{L-L}\left( \gamma \right) $ at the
limit $\alpha =1$. In intermediate value of $\alpha $\ and $\gamma $, $\tilde{e}%
\left( \gamma ,\alpha \right) $ deviates with a maximum relative error of $%
0.03$ from the numerical result at $ \gamma \approx 2.5,\alpha \approx
0.9 $. In Fig.\ref{comparee}, we exhibit the result of exact
numerical result of $e\left( \gamma ,\alpha \right) $ compared with the
fitting formulas $\tilde{e}\left( \gamma ,\alpha \right) $ for various
interaction strength $\gamma $ and the fraction of Bosons $\alpha $. Clearly
the fitting formulas represent quite well the Betha-Ansatz result for the
whole range of interaction and arbitrary fraction of bosonic atoms in the
mixture. These formulas are then adopted in the following solution of the
KSEs equations.

From $\varepsilon ^{\hom }\left( n,\alpha ,\gamma \right) $, the ground
state Bose and Fermi chemical potentials can be obtained as%
\begin{eqnarray}
\mu _{B}^{\hom }\left( n,\alpha ,\gamma \right) &=&\frac{\hbar ^{2}n^{2}}{2m}%
f_{B}\left( \gamma ,\alpha \right) \text{,}  \notag \\
\mu _{F}^{\hom }\left( n,\alpha ,\gamma \right) &=&\frac{\hbar ^{2}n^{2}}{2m}%
f_{F}\left( \gamma ,\alpha \right) \text{,}  \label{chemical}
\end{eqnarray}%
where%
\begin{eqnarray}
f_{B}\left( \gamma ,\alpha \right) &=&3e-\gamma \frac{\partial e}{\partial
\gamma }+\left( 1-\alpha \right) \frac{\partial e}{\partial \alpha },  \notag
\\
f_{F}\left( \gamma ,\alpha \right) &=&3e-\gamma \frac{\partial e}{\partial
\gamma }-\alpha \frac{\partial e}{\partial \alpha }\text{,}  \label{f}
\end{eqnarray}%
with $f_{B}\left(0,\alpha \right)=0,f_{F}\left( 0 ,\alpha \right)=\pi^2 \left( 1-\alpha \right) ^2$, and
$f_{B}\left( + \infty,\alpha \right)= f_{F}\left( + \infty ,\alpha \right)=\pi^2$.


\section{Numerical results}

\begin{figure}[tbp]
\includegraphics[width=0.45\textwidth]{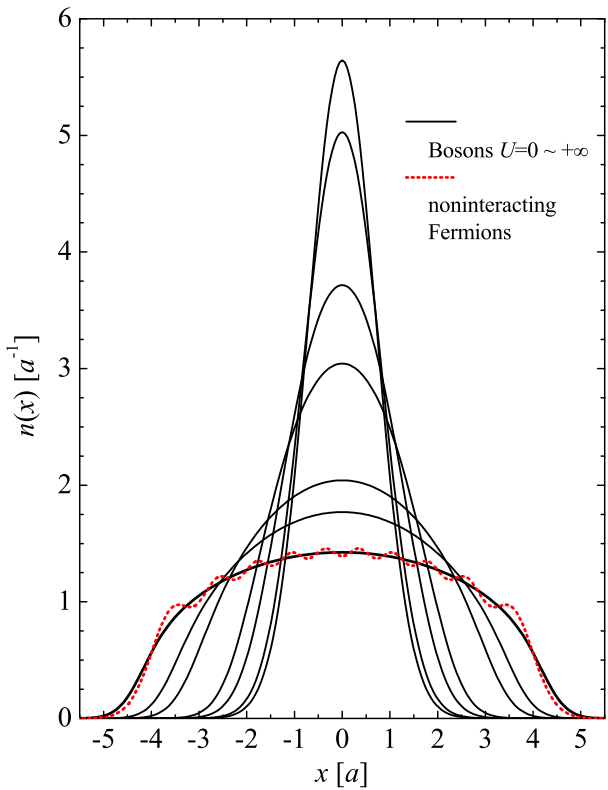}
\caption{(Color Online) Density distribution of $N_F=10$ noninteracting Fermions (red dotted
line) and the density distributions of $N_B=10$ Bosons for different interaction 
parameter $U$. The six black solid lines from top to bottom are respectively
for Bosons with $U=0,0.1,0.5,1,5,10,+\infty$.}
\label{10NBdensity}
\end{figure}

\begin{figure}[tbp]
\includegraphics[width=0.45\textwidth]{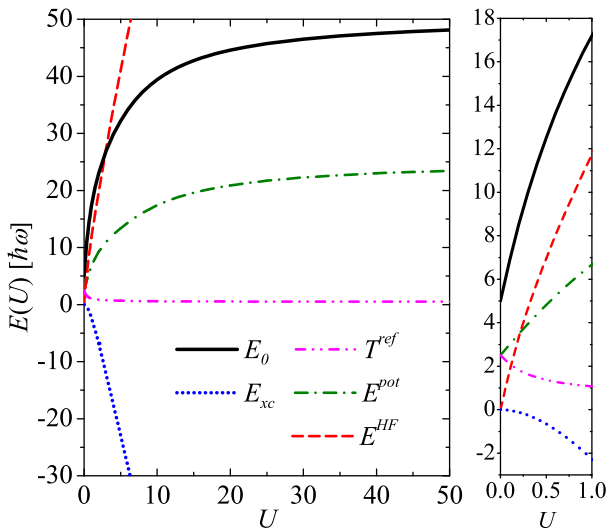}
\caption{(Color Online) Evolution of energies of $N_{B}=10$ Bosons with increasing interaction parameter $U$. 
Contributions to the ground state energy $E_0$ include: 
kinetic energy $T^{ref}$, external potential energy $E_{pot}$, Hatree-Fock
energy $E_{HF}$, exchange correlation energy $E_{xc}$. Right Panel: Details in the
mean-field regime. }
\label{10NBenergy}
\end{figure}

Inserting (\ref{chemical}) into the KSEs (\ref{KSB}) and (\ref{KSF}), and
assuming Bosons and Fermions suffer from the same harmonic external
potentials $V_{B}(x)=V_{F}(x)=m\omega ^{2}x^{2}/2$, with $\omega $\ is
frequency, we can get the ground state density profiles of each component by
solving the KSEs together with the constraint (\ref{n}) by means of
numerical iteration. The ground state energy follows immediately from
(\ref{E0c}). Here we introduce the length unit $a=\sqrt{\hbar /m\omega }$ 
and a dimensionless interacting parameter $U=g/a\hbar \omega $ such 
that the space dependent Lieb-Liniger parameter is expressed as 
$\gamma (x)=U/ an(x) $. Before going into the details of the DFT result, 
we first discuss the KSEs for some limiting cases.

When there is no interactions in the
mixture, $U=0$, KSEs correctly reduce to the equations for noninteracting Bosons and
noninteracting Fermions in the harmonic trap. 
The densities of bosonic and fermionic components are respectively
\begin{eqnarray}
n_{B}\left( x\right) &=&\frac{N_{B}}{a\sqrt{\pi }}\exp \left(
-x^{2}/a^{2}\right) ,  \label{bose density} \\
n_{F}\left( x\right) &=&\frac{1}{a\sqrt{\pi }}\exp \left(
-x^{2}/a^{2}\right) \sum_{l=0}^{N_{F}-1}\frac{H_{l}^{2}\left( x/a\right) }{%
2^{l}l!},  \label{fermion density}
\end{eqnarray}%
where $H_{l}\left( x\right) $ is the Hermite polynomials. Here the
noninteracting Bose density profile (see the topmost black line in Fig. %
\ref{10NBdensity}) is a standard Gaussian-like shape and Fermi density
profile (see the red dotted line in Fig. \ref{10NBdensity}) is characterized
by a half ellipse-like shape with $N_{F}$ oscillations. The ground state
energies of these two components are%
\begin{eqnarray}
E_{0B} &=&\frac{N_{B}}{2}\hbar \omega ,  \label{boson energy} \\
E_{0F} &=&\sum_{l=0}^{N_{F}-1}\left( l+\frac{1}{2}\right) \hbar \omega
\label{fermion energy}
\end{eqnarray}%
and the total ground state energy is $E_{0}=E_{0B}+E_{0F}$. 

When the interaction is weak, neglecting $E_{xc}$ in (\ref{E0}), the KSEs
reduce to our familiar mean-field formulas%
\begin{equation*}
\left( -\frac{\hbar ^{2}}{2m}\frac{d^{2}}{dx^{2}}+\frac{1}{2} m \omega^2 x^2
+g\left(n_{B}+n_{F}\right)\right) \phi =\epsilon \phi ,
\end{equation*}%
\begin{equation}
\left( -\frac{\hbar ^{2}}{2m}\frac{d^{2}}{dx^{2}}+\frac{1}{2} m \omega^2 x^2
+gn_{B}\right) \psi _{j}=\eta _{j}\psi _{j}.  \label{mean field}
\end{equation}
For bosons it is nothing but the Gross-Pitaevskii equation
for dilute gas. The equation for fermions, on the other hand, 
reminds us the superfluid theory of the mixture of bosons and 
paired BCS states of the two-component fermions where 
only one Fermionic orbital is considered \cite{Adhikari2010}.

When the interaction is strong, for system of large atom numbers $N_B,N_F \gg 1$, 
one can safely use the Thomas-Fermi
approximation (TFA), i.e., the kinetic energies $T_{B}^{ref}$ and $%
T_{F}^{ref}$ in the energy functional (\ref{E0}) are approximated to zero
and $\int n\left( x\right) \kappa _{s}^{\hom }\left( x\right) dx$,
respectively. Minimizing $E_{0}$ directly with respect to $n_{B}\left(
x\right) $ and $n_{F}\left( x\right) $, we get the TFA formulas
\begin{eqnarray}
\frac{1}{2} m \omega^2 x^2 +\mu _{B}^{\hom }\left( \left[ n_{B},n_{F}\right]
;x\right) &=&\mu _{B}^{0},  \notag \\
\frac{1}{2} m \omega^2 x^2 +\mu _{F}^{\hom }\left( \left[ n_{B},n_{F}\right]
;x\right) &=&\mu _{F}^{0},  \label{TFA}
\end{eqnarray}%
where $\mu _{B}^{0}$ and $\mu _{F}^{0}$ are constants fixed by the
normalization conditions $\int n_{B}\left( x\right) dx=N_{B}$ and $\int
n_{F}\left( x\right) dx=N_{F}$. Eqs. (\ref{TFA}) are explained as the LDA of
the chemical potentials at point $x$ in \cite{Imambekov} and have 
been used extensively \cite{Guan2008,Xiangguo Yin}.
That means,  in slowly varying external harmonic trap chemical potentials 
at point $x$ are related to those in the trap center $x=0$ ($\mu _{B}^{0}$ and $\mu _{F}^{0}$).

When repulsive interactions are infinitely strong, $\mu _{B}^{\hom }=\mu _{F}^{\hom }
=\hbar ^{2}\pi ^{2}n^{2}/2m$, (\ref{TFA}) reduces to a single equation%
\begin{equation}
\frac{1}{2}m\omega ^{2}x^{2}+\frac{\hbar ^{2}\pi ^{2}}{2m}n^{2}\left(
x\right) =\mu ^{0},  \label{TFAn}
\end{equation}%
with $\mu ^{0}$ is decided by $\int n\left( x\right) dx=N$. This gives us
the explicit result for total density distribution
\begin{equation}
n\left( x\right) =\frac{\sqrt{2N-x^{2}/a^{2}}}{\pi a},
\end{equation}%
and ground state energy 
\begin{equation}
E_0=\frac{N^2}{2} \hbar \omega. 
\end{equation}
We see that they are exactly the density distribution and energy
of $N$ free fermions in a harmonic trap. Equation (\ref{TFAn}), however, gives nothing
about the densities of Bose and Fermi components. The method here is insufficient for the
infinitely strong interaction. We may, on the other hand, resort to the Bose-Fermi mapping
method \cite{Girardeau2007,Bess Fang2009} which gives the Bose and Fermi density profiles as
\begin{equation}
n_{B,F}\left( x\right)  = \frac{N_{B,F}}{N\sqrt{\pi }}\exp \left(
-x^{2}/a^{2}\right) \sum_{n=0}^{N-1}\frac{H_{n}^{2}\left( x\right) }{2^{n}n!}. \label{nbf}
\end{equation}
The two components are nondemixing in agreement with the generalized Bethe ansatz wave function
\cite{Bess Fang2011}, 

The DFT results are summarized in Figs. 2-8. First, for a pure
bosonic system, equation (\ref{KSB}) is just the generalized
Gross-Pitaevskii equation appeared in Refs. \cite{Kim2003,DFT for 1D bose}. We show
the density profiles for $N_{B}=10$ bosons in Fig. \ref{10NBdensity} 
for the cases of $U=0,0.1,0.5,1,5,10$ and $+\infty $, respectively. With the increasing of
$U$, the density profiles vary from a standard Gaussian-like to a
non-oscillating half-ellipse shape. Comparing the density profiles of Bosons at $U=+\infty $ 
and the noninteracting fermions, we find that they
match each other quite well except the density oscillations. The
results mean the density distribution of Bosons with infinitely
strong repulsive delta interaction is basically the same as that of a
noninteracting Fermi gas which is consistent with the theory of Bose-Fermi
mapping theory \cite{Girardeau1960}. 
Our theory fails to reproduce the density oscillation
due to the impenetrable property of 1D system with strong interaction because
we adopt one single functional orbital $\phi(x)$ for the 1D Bose liquid. 
The exact oscillations reflecting the structure of the occupied orbitals should
be quested from the real wave function such as by the exact
diagonalization method \cite{Deuretzbacher2007}.
In the limit of large particle number the differences between the oscillating 
and non-oscillating profiles become imperceptible. 

The ground state energy evolution as a function of $U$ is illustrated in Fig. \ref%
{10NBenergy}. We can see that with the increase of $U$, the kinetic energy $%
T^{ref}$ decreases slowly indicating that the
interaction restrains the movement of atoms. The external potential energy
$E_{pot}$ increases as a result of wider
and wider occupied regime of the trap. Both of these two energies evolve to
constant energies. The Hatree-Fock energy $E_{HF}$ increases
almost linearly while the exchange correlation energy $E_{xc}$ decreases in the
whole interaction regime. These two terms play more and more important roles
in the DFT theory for stronger interaction and they approximately cancel each other.
All these energies contribute to the total energy $E_{0}$, which starting from 
the noninteracting value $5\hbar \omega $ approaches the strongly interacting
limit $50\hbar \omega $. For $U<0.9$, the exchange correlation energy is much less 
than the total energy, $\left\vert E_{xc}/E_{0}\right\vert <0.1$, which can be seen as the
effective regime of mean field theory. For a TG gas with $U=+\infty $ and a chemical potential
$\mu _{B}^{\hom } =\hbar ^{2}\pi ^{2}n ^{2}/2m$, numerically 
solving equation (\ref{KSB}) gives $E_{0}=50.5024\hbar \omega $ which lies
slightly above $50\hbar \omega $ because the introduced Bose
functional orbital $\phi (x)$ is only an assistant variational function instead of
the true wave function of the interacting Bose system.

\begin{figure}[tbp]
\includegraphics[width=0.45\textwidth]{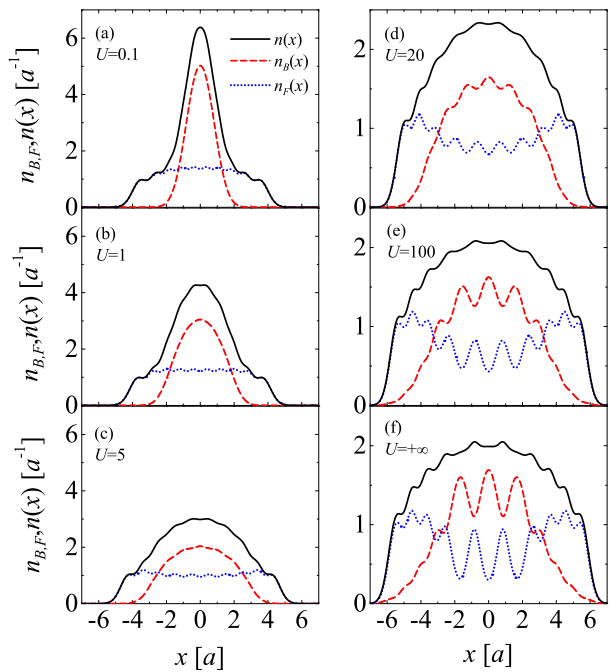}
\caption{(Color Online) Density distributions of a mixture of $N_B=10$ bosons and $N_F=10$ fermions 
for different interaction parameter $U=0.1,1,5$ (left) and $U=20,100,+ \infty$ (right). }
\label{10NB10NFdensity}
\end{figure}

\begin{figure}[tbp]
\includegraphics[width=0.45\textwidth]{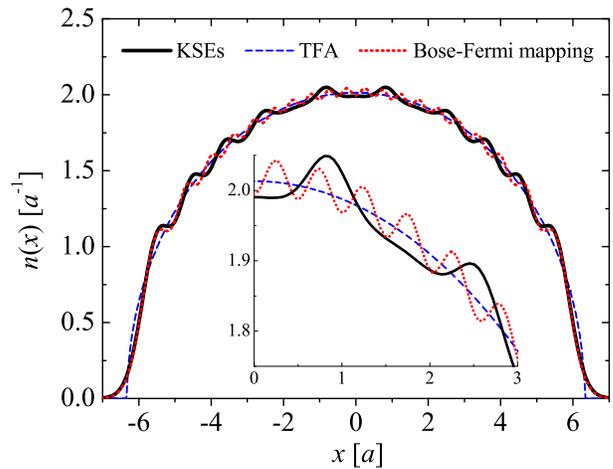}
\caption{(Color Online) The total density profiles $n$ as a function of $x$ for a $N_B=N_F=10$
mixture with $U=+\infty $. The result of KSEs (black solid line), TFA (blue
dashed line) and Bose-Fermi Mapping (red dotted line) are
compared and the inset shows a zoom into the structure of the ocsillations. }
\label{10NB10NFTGdensity}
\end{figure}

\begin{figure}[tbp]
\includegraphics[width=0.45\textwidth]{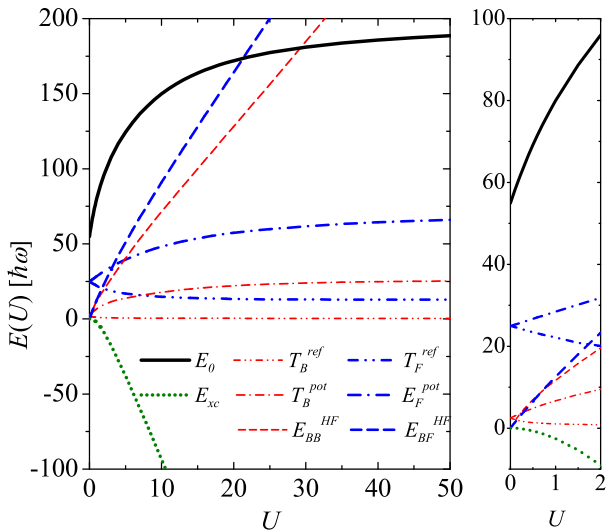}
\caption{(Color Online) Ground state energies as a function of $U$ for mixture of 
$N_{B}=N_{F}=10$. Contributions to the ground state energy $E_0$ are 
similar to those in Fig. 3. All terms but the exchange correlation energy originate
from bosons and fermions. Right Panel: Details in the mean-field regime.}
\label{10NB10NFenergy}
\end{figure}

We now turn to illustrate the main result of a mixture of $N_{B}=10$ bosons and 
$N_{F}=10$ fermions. The densities of non-interacting mixture (\ref{bose density})
and (\ref{fermion density}) are taken as the starting point of the iteration of KSEs 
(\ref{KSB},\ref{KSF}) for a small interaction parameter (e.g. $U=0.1$). The eigenvalues $\epsilon, \eta_j$
and functional orbitals $\phi,\psi_j$ are found by iterating to the desired degree 
of accuracy. The new densities are initial densities for the next iteration for a 
larger interaction parameter, and so on. The density profiles for different $U$ are displayed in 
Fig. \ref{10NB10NFdensity}. It shows that with increasing
$U$, the peak of the total density $n(x)$ decreases monotonically and
atoms tend to occupy wider regime. The density of Fermi component 
changes smoothly in amplitude, while Bose component becomes more and 
more flat and ripples begin to appear for stronger interaction. 
At weak interaction, $U=0.1,1$ as in Fig. \ref{10NB10NFdensity} (a)
and (b), both Bosons and Fermions are located in the center of the trap. For an 
intermediate interaction strength, e.g. $U=5$ as in Fig. \ref{10NB10NFdensity} (c), some
fermions are excluded from the center of the trap while Bosons are held
mainly in the center. We notice that oscillations emerge in the Bose density curves
reflecting the strong correlation with Fermions. When $U$
becomes further stronger, $U=20,100$ as in Fig. \ref{10NB10NFdensity}(d) and (e), more
Fermions are repelled out from the center and a clear signature of phase separation 
of bosons and fermions is seen in the figures. High density of discrete bosons 
are surrounded by fermions, which nevertheless still have chance to 
squeeze between the opening space of bosons. The total density profile approaches 
half ellipse-like for $U=+ \infty$ as shown in Fig. \ref{10NB10NFdensity}(f).
We may have a close inspection of the case of infinitely strong interaction.
In Fig. \ref{10NB10NFTGdensity}, the DFT result of the total density is compared with those from TFA and 
Bose-Fermi mapping. It is clear that the agreement is fairly good except tiny 
difference in the number, position and amplitude of the
oscillations which are enlarged in the inset of Fig. \ref%
{10NB10NFTGdensity}. Again for large atom number the differences 
between the oscillating and non-oscillating curves is unperceivable.

\begin{figure}[tbp]
\includegraphics[width=0.45\textwidth]{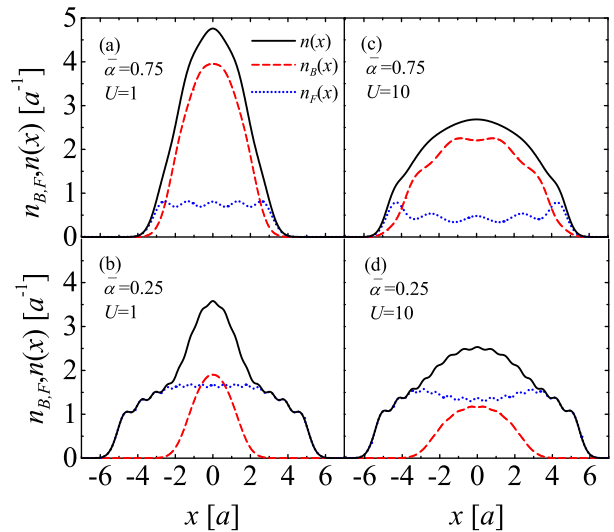}
\caption{(Color Online) Density distributions of mixture with $N=20$ for different mean fraction of 
bosons $\bar \alpha =N_B/N$ and different interaction parameter $U$.}
\label{NBNFdensity}
\end{figure}

Fig. \ref{10NB10NFenergy} describes the evolution of all contributed energy terms in (%
\ref{E0}) as a function of $U$. The line types denote the same energy terms as in 
Fig. \ref{10NBenergy} except that here the kinetic energy and
external potential energy respectively include two terms relating to
Bosons (in red) and Fermions (in blue). The trend of these lines
resemble those in Fig. \ref{10NBenergy} for the same reason. Especially the 
exchange correlation energy here contest with two Hatree-Fock terms
representing the mean field energy of boson-boson and boson-fermion interaction respectively. The total
ground state energy evolves from $55\hbar \omega $, the energy of $10$ ideal bosons
and $10$ ideal fermions, to $200\hbar \omega $,
the energy of $20$ fully fermionized atoms according
to (\ref{boson energy}) and (\ref{fermion energy}). In the parameter 
range of $0<U<2$, $\left\vert E_{xc}/E_{0}\right\vert <0.1$, mean field theory is 
regarded effective.
At $U=+\infty $, we numerically obtain an upper bound for the ground state 
energy $E_{0}=200.1364\hbar \omega $ which is very close to the exact result of 
full fermionization $E_{0}=200\hbar \omega $.

There are some two contradicting predictions for the spatial structure of the components 
densities of trapped TG mixture. Based on the exact result of Bethe ansatz and TFA formulas 
(\ref{TFA}) (where they called it LDA), Ref. \cite{Imambekov} gives a phase 
separation result of the two components for a strong but finite interaction. The 
thermodynamical Bethe ansatz (TBA) at finite temperature \cite{Xiangguo Yin} 
tends to support this scenario.
The Bose-Fermi mapping methods in Ref. \cite{Girardeau2007} and 
\cite{Bess Fang2009} is proposed in TG limit of infinitely strong interaction 
and the results show that the component distributions of the $N_{B}=N_{F} $ 
mixture are completely the same according to (\ref{nbf}) therefore display no demixing. But the authors of 
\cite{Bess Fang2009} have noticed that the ground state given in this way is highly
degenerate. They subsequently used a generalized Betha ansatz wave function,
in which the 'orbital' part of the wave function is essentially replaced by a Slater 
determinant of single-particle Schr\"{o}dinger equation in the trap potential,
to give a nondemixing result at finite large interactions. They further tested this 
nondemixing result with numerical DMRG simulations for a lattice model of 
$N_{B}=N_{F}=2$ mixture \cite{Bess Fang2011}. We observe, however, obvious signature 
of phase separation in Fig. 4 of \cite{Bess Fang2011} for relatively large 
interaction $U=100$. The intrinsic nature of the phase 
separation and nondemixing in TG limit originates from the Bethe ansatz and Bose-Fermi 
mapping techniques respectively. We expect experimental verification of the nature of
spatial configuration about trapped ultracold atomic mixtures.

\begin{figure}[tbp]
\includegraphics[width=0.45\textwidth]{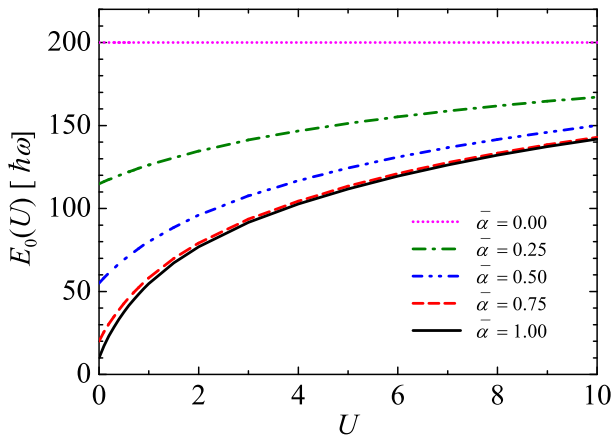}
\caption{Ground state energy $E_{0}$ as a function of $U$ for different fraction of 
bosons $\bar \alpha$ in the mixture with total atom number $N=20$.}
\label{NBNFenergy}
\end{figure}

Finally we discuss the effect of another system parameter, i.e. mean value of 
the fraction of bosonic atoms number $\bar \alpha=N_B/N$, on the density profiles. 
Fig. \ref{NBNFdensity} shows the density of each component and the 
total density of $N=20$ atoms in the mixture with 
$\bar \alpha =0.25, 0.75$ and interaction parameters $U=1,10$. Fig. \ref{NBNFenergy}
compares their energies. Bosons will dominate the total density profile when
more bosons are put into the mixture for weak as well as strong interaction. 
When more fermions are prepared in the gas, bimodal distribution is clearly seen the 
total density where bosonic Gaussian shape is superimposed onto the fermionic 
shell-like structure. In the strong interaction limit the total density approaches to the 
typical half-ellipse no matter how many bosons or fermions are involved in the mixture.
The number of fermions contribute to the ground state energy more effectively in the weak 
interaction case. This situation changes for strong interaction where the energies for all
values of $\bar \alpha$ approximate to the limit value of the fully fermionization of 
the system.

\section{Conclusion}

In conclusion, using the DFT we study the ground state energy and
density distribution of the Bose-Fermi mixture in a quasi 1D
harmonic trap. Based on the Bethe ansatz solution for the mixture, 
we managed to obtain a fitting formula for the function $e(\gamma,\alpha)$ for the 
ground state energy of homogeneous system. The KSEs are obtained from 
the variational minimization of the energy functional of trapped mixture
with respect to the densities of bose and fermi components.
We found that when the interaction between the atoms varies
from zero to positive infinitely, the ground state energy of the mixture would 
evolve to the constants of the noninteracting Fermions and the total
density approached a half ellipse profile. More and more fermions are repelled 
out of the trap center, while bosons occupy the central region. Phase separation
of boson and fermion components occurs for strong interaction in agreement with the
the result Bethe ansatz method plus LDA. The calculation here applies equally
to the pure bosonic case, different fraction of bosons, as well as in the TG limit. 
Our DFT theory is also suitable for
mixtures in optical lattice and could be extended to study the dynamical
and thermodynamic phenomena of the mixture.

\begin{acknowledgments}
This work is supported by the NSF of China under Grant Nos. 11104171 and
11074153, the National Basic Research Program of China (973 Program) under
Grant Nos. 2010CB923103, 2011CB921601, the NSF of Shanxi Province, and 
the Program for New Century Excellent
Talents in University (NCET). We thank Gao Xianlong for helpful discussions.

\end{acknowledgments}

\end{document}